\documentclass[11pt]{article}
\usepackage{moriond,epsfig}

\def\ra{$\rightarrow$}

\def\g{\gamma}
\def\ww{$W^+W^-$ }
\def\wwg{$W^+W^-\gamma$ }
\def\swwg{\sigma_{WW\gamma}}
\def\ee{$e^+e^-$}
\def\nngg{$\nu\bar\nu\gamma\gamma$ }
\def\vvgg{$\nu_e\bar\nu_e\gamma\gamma$ }
\def\a0{a_0/\Lambda^2}
\def\ac{a_c/\Lambda^2}
\def\an{a_n/\Lambda^2}

\def\bea{\begin{eqnarray}}
\def\eea{\end{eqnarray}}

\begin{document}
\vspace*{4cm}
\title{QUARTIC GAUGE BOSON COUPLINGS RESULTS AT LEP}

\author{M. MUSY }

\address{CERN}

\maketitle\abstracts{The study of charged and neutral boson vertices
  has been performed in different production channels at the LEP
  experiments. Decay rates and kinematic properties of these events
  are exploited to set constraints on the corresponding gauge
  couplings.}

\section{Introduction}
\begin{figure}[htb]
  \begin{center}
    \begin{tabular}{ccc}
\includegraphics*[width=.32\textwidth]{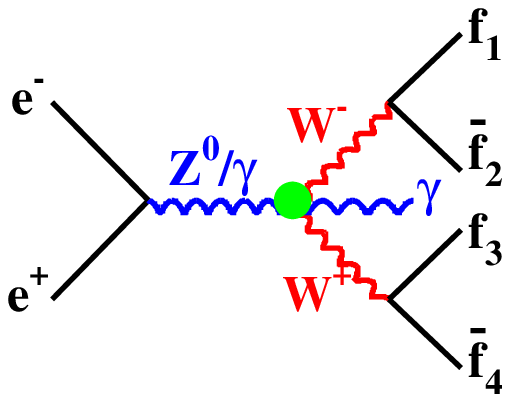} &
\includegraphics*[width=.24\textwidth]{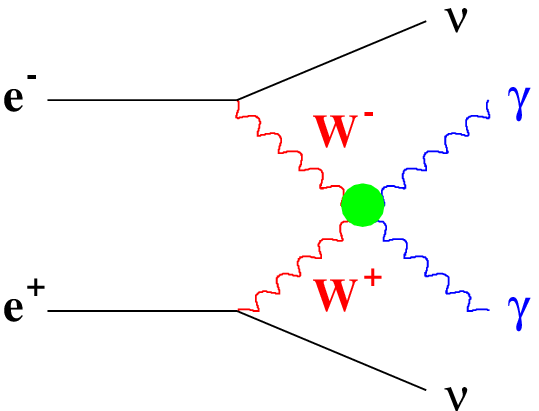} &
\includegraphics*[width=.3\textwidth]{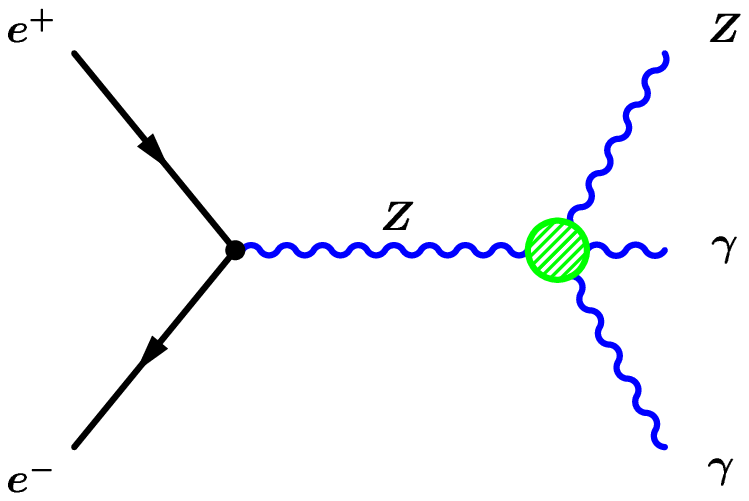} \\
    \end{tabular}
    \label{fig1}
\caption{Feynman diagrams containing a four boson vertex leading to 
  the (a) \wwg, (b) \nngg, and (c) Z$\g\g$ final states.}
  \end{center}
\end{figure}

The study of quartic gauge boson couplings (QGCs) has become possible
due to the recent theoretical developments on this topic. The presence
of QGCs affects the data collected for different final states at LEP.
The continous increase of the centre-of-mass energy in \ee collisions
allowed for the precise measurement of the W-pair cross section and
couplings. Now, also the study of radiative W-pair events, \ee\ra\wwg,
has become possible.  The Standard Model (SM) predicts the existence
of quartic gauge boson couplings leading to \wwg production via
$s$-channel exchange of a $\g$ or a Z boson as shown in
Fig.~\ref{fig1}a.  The contribution of these two quartic Feynman
diagrams with respect to the other competing diagrams, mainly initial
state radiation, is negligible at the LEP centre of mass energies.
Nonetheless, the process leading to the \wwg final state can be
sensitive to anomalous contributions to the SM quartic vertices
\wwg$\g$ and \ww Z$\g$.

The theoretical framework of Ref.[~\cite{anja}] is used for the
parametrisation of such anomalous couplings.

The existence of Anomalous QGCs would also affect the \ee\ra\vvgg
process via the \ww fusion Feynman diagram containing the \wwg$\g$
vertex~\cite{anja2} (see Figure~\ref{fig1}b). In the SM the reaction
\ee\ra\nngg proceeds predominantly through $s$-channel Z exchange and
$t$-channel W exchange, with the two photons coming from initial state
radiation, whereas the SM contribution from the \ww fusion is again
negligible at LEP. AQGCs would enhance the \nngg production rate,
especially for the hard tail of the photon energy distribution and for
photons produced at large angles with respect to the beam direction.

Finally, the Z$\g\g$ production (see Fig.~\ref{fig1}c) can also be
exploited to derive limits on AQGCs, as it will be discussed in the
next paragraph.

\section{QGCs Analyses at LEP}
Table 1 shows the different data sets used for the AQGCs
analyses at LEP.

\begin{table}[htb]
\begin{center}
\begin{tabular}{|c|c|c|c|}
\hline
 &Aleph &L3 &Opal \\
\hline
\wwg & -- &189-202 GeV &189 GeV  \\ 
\nngg& 189-202 GeV &183-202 GeV &189 GeV  \\ 
Z$\g\g$ & -- &130-202 GeV &130-208 GeV  \\
\hline
\end{tabular}
\end{center}
\label{tab1}
\caption{Data sets used for the  AQGCs analyses.}
\end{table}

The quartic couplings corresponding to reactions where two or more 
{\it charged} intermediate vector bosons are involved, are those 
leading to the \wwg and \nngg final states.

In the \wwg analysis, there are many Feyman diagrams leading at the
$\cal{O}$$(\alpha)$ leading to the ffff+$\g$ final state (e.g. 142
diagrams only for \ee\ra\wwg\ra$u\bar d e^- \bar\nu_e\g$). Anomalous
contributions to the SM quartic diagrams manifest themselves through
the hardening of the photon spectrum, which has the highest
sensitivity with respect to the other kinematic variables in \wwg
events, and it is therefore used to extract the limits.

In the WW semileptonic and hadronic samples, an isolated photon 
is selected assuming different definitions for the phase space. The L3
and Opal experiments require cuts on the polar and separation angle
between the photon and the charged fermions in the final state, and on 
the $f\bar f$ invariant mass (for Opal).
Table~\ref{tab2} summarises the phase space definitions assumed.

\begin{table}[htb]
\begin{center}
\begin{tabular}{|c|c|}
\hline
 L3 &Opal \\
\hline
$E_\g>5$ GeV &$E_\g>10$ GeV\\
$|\cos\theta_\g|<0.94$&$|\cos\theta_\g|<0.9$\\
$\cos\theta_{f\g}<0.94$&$\cos\theta_{f\g}<0.9$\\
--&min$M_{ff'}>73$ GeV\\
\hline
\end{tabular}
\end{center}
\label{tab2}
\caption{Phase space definition for the photon in \wwg events.}
\end{table}

%figura cross section
\begin{figure}[htb]
  \begin{center}
\includegraphics*[width=.82\textwidth]{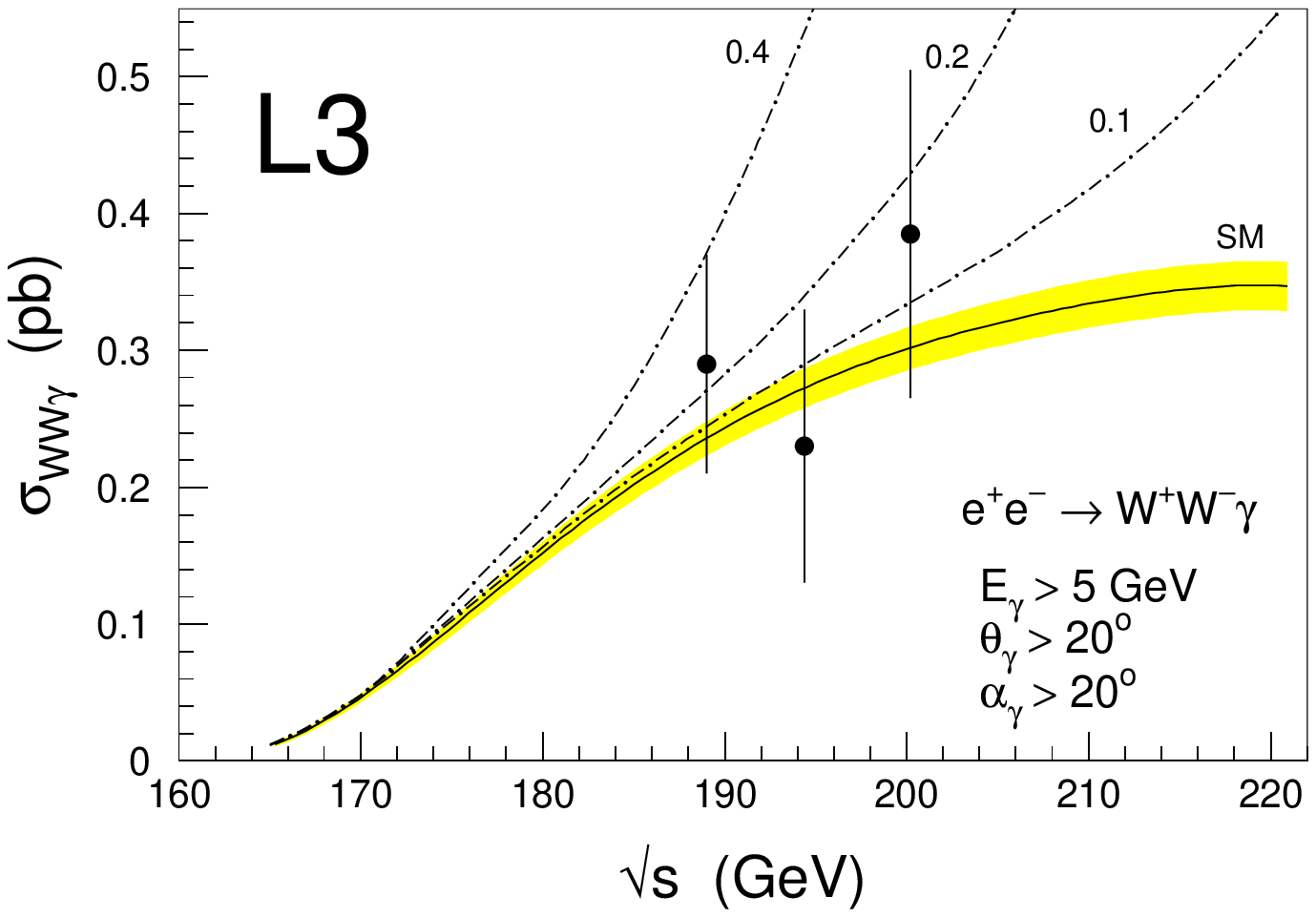}    
\vspace{-1.5cm}
    \caption{Measured cross section for the process \ee\ra\wwg. }
    \label{fig2}
  \end{center}
\end{figure}

\newpage
The resulting cross sections for the L3 experiment are: 
\bea
\swwg(188.6 {\rm ~GeV})&= 0.29\pm 0.08\pm 0.02 {\rm ~pb} \nonumber\\
\swwg(194.4 {\rm ~GeV})&= 0.23\pm 0.10\pm 0.02 {\rm ~pb} \nonumber\\
\swwg(200.2 {\rm ~GeV})&= 0.39\pm 0.12\pm 0.02 {\rm ~pb} \nonumber
\eea 
while for Opal, the
measurement at 189 GeV gives $\swwg=0.136\pm 0.037\pm 0.008$ pb, 
where the first error is statistic and the second systematic.\\
All the obtained results are compatible with the SM expectations.\\
Figure \ref{fig2} shows the L3 measured cross section as a fuction 
of $\sqrt{s}$. The expectation band corresponds to the 5\% uncertainty
on the theoretical prediction from the {\tt EEWWG}~\cite{anja}
Monte Carlo. The three dashed lines correspond to the cross section 
for non vanishing values of $\a0$, $\ac$, and $\an$\ (in GeV$^{-2}$ units).

\begin{figure}[htb]
\begin{center}
    \begin{tabular}{c}
      \hspace*{-1.4cm}
      \epsfig{file=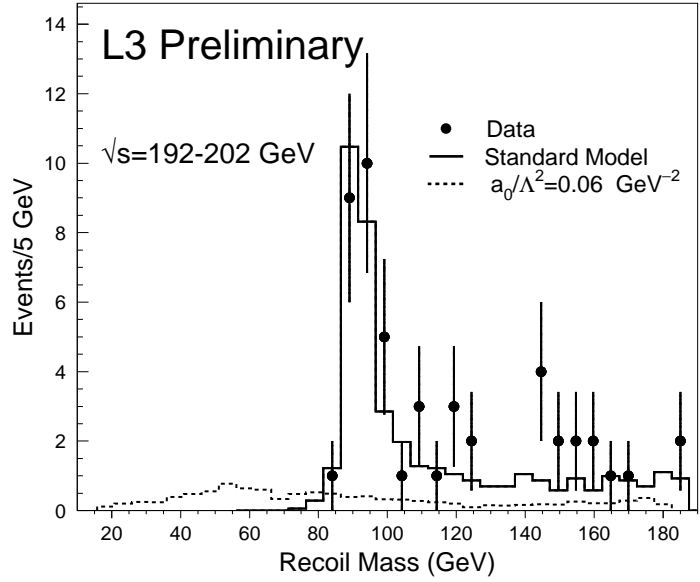, width=10cm}\\
      \hspace*{-1.cm}
      \epsfig{file=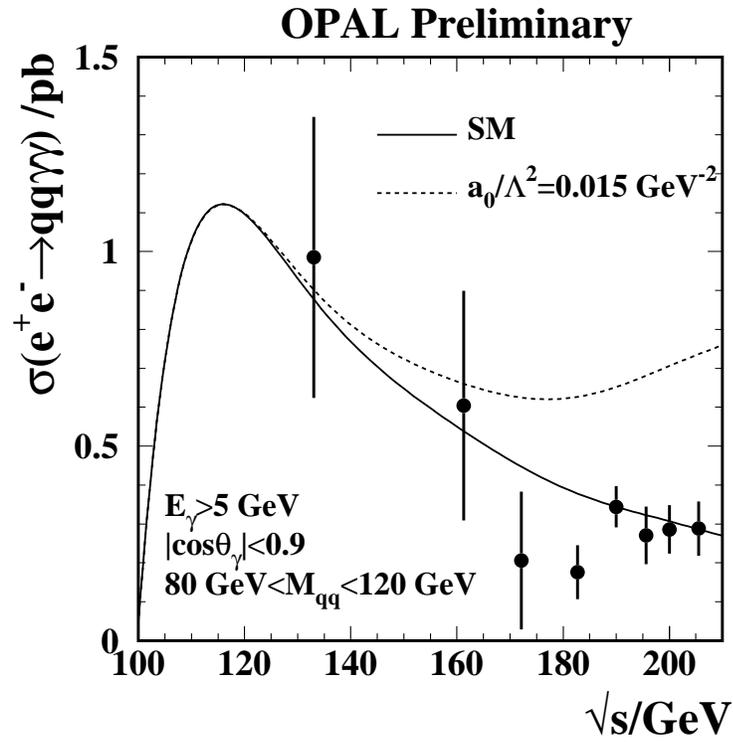, width=11cm}\\
    \end{tabular}
\end{center}  
  \caption{Cross section for the 
    process \ee\ra Z$\g\g$ as a function of $\sqrt{s}$.}
  \label{fig3}
\end{figure}

The acoplanar multi-photon channell is also sensitive to AQGCs. For a
large part of the AQGC signal, the two photons recoil mass is less
than the mass of the Z, while almost no SM contribution is expected in
this kinematic region.  Figure~\ref{fig3} shows the differential
distribution in the specified centre-of-mass range. The {\tt
  EENUNUGGANO}~\cite{anja} program does not describe the effects of the
SM $s$-channel Z exchange diagrams and the interference terms among
these diagrams and the \ww fusion diagram containing the \wwg vertex.
Therefore, the dashed line corresponding to the signal is not reliable
ove $M_Z$, and a recoil mass cut is applied to suppress  the SM contribution.

\newpage 
The information from the \nngg total cross section, together
with the information derived from the shape and normalisation of the
photon spectra in \wwg events, produces the following ADL combined
limits on {\it charged} AQGCs using the data set of Table~1
(first two lines):
\bea
-0.022 {\rm ~GeV}^{-2} &< \a0 < ~0.021 {\rm ~GeV}^{-2} \nonumber\\
-0.043 {\rm ~GeV}^{-2} &< \ac < ~0.058 {\rm ~GeV}^{-2} \nonumber\\
-0.022 {\rm ~GeV}^{-2} &< \an < ~0.020 {\rm ~GeV}^{-2}, \nonumber
\eea
in good agreement with the SM expectation of zero for each coupling.\\

The Feynman diagram of Fig.\ref{fig1}c involves only {\it neutral}
bosons, and it is not present in the SM. Besides, as it has been
recently indicated~\cite{boh}, under more general assumptions the
quartic couplings in the neutral sector, still named $\a0$, and $\ac$\ 
in Ref.[~\cite{anja}], can be regarded as independent from those in the
charged sector. For this reason, on the experiimental side, the
convention was used to keep apart the measurements in these channels
not performing any combinations, in the wait for a unified theoretical
approach.

To obtain limits on AQGCs in the neutral sector, the hadronic
Z$\g\g$\ra qq$\g\g$ events are used. The signal consists in an
enhancement in the rate of high energy photons, while the background
mainly comes from doubly radiative returns to the Z. Figure \ref{fig3}
shows the Z$\g\g$ cross section as predicted by the {\tt
  EEZGG}~\cite{anja2} program~\footnote{The results on \wwg and
  Z$\g\g$ cross section will be reproduced on the base of a common
  definition for the phase space of the photon.}. The same program is
used to model AQGCs and to fit these couplings to the observed data
distributions 
of $P_{T\g 1}$ (for L3), $E_\g$ and max$|\cos\theta_{\g i}|$ (for Opal).\\
The following LEP combined bounds are derived~\cite{ichep}: 
\bea
-0.0048 {\rm ~GeV}^{-2} &< \a0 < ~0.0056 {\rm ~GeV}^{-2} \nonumber\\
-0.0052 {\rm ~GeV}^{-2} &< \ac < ~0.0099 {\rm ~GeV}^{-2} \nonumber 
\eea
All experimental observations are compatible with the SM predictions.

\section{Conclusions}
The measurements of rare processes as Z$\g\g$, \wwg and \nngg
constitute an important test of the SM. Preliminary results on AQGCs
were presented here, including new results in the \wwg channel up to
the centre-of-mass energy of 202 GeV with a significant increase in
the final precision.

\end{document}